\documentstyle[psfig]{l-aa}
\def\jref#1 #2 #3 #4 {{\par\noindent \hangindent=2em \hangafter=1
      \advance \rightskip by 0em #1, {\it#2}, {\bf#3}, #4.\par}}
\def\rref#1{{\par\noindent \hangindent=2em \hangafter=1
      \advance \rightskip by 0em #1.\par}}

\def\COBE{{\sl COBE\/}}
\def\wisk#1{\ifmmode{#1}\else{$#1$}\fi}

\def\etal   {et~al.\,}

\def\deg{\ifmmode^{\circ}\else$^{\circ}$\fi} 
\def\min{\ifmmode^{\prime}\;\else$^{\prime}\;$\fi}
\def\sec{\ifmmode^{\prime\prime}\;\else$^{\prime\prime}\;$\fi}

\def\dn{\ifmmode{\Delta\nu{_d} }\else{$\Delta\nu_{d}$ }\fi}
\def\dt{\ifmmode{\Delta t{_d} }\else{$\Delta t_{d}$ }\fi}

\def\lsim{\,\lower2truept\hbox{${< \atop\hbox{\raise4truept\hbox{$\sim$}}}$}\,}
\def\gsim{\,\lower2truept\hbox{${> \atop\hbox{\raise4truept\hbox{$\sim$}}}$}\,}

\title{The Planck-LFI instrument: analysis of the $1/f$ noise and
implications for the scanning strategy}

\author{{D. Maino}\inst{1} \and {C. Burigana}\inst{2} 
\and {M. Maltoni}\inst{2,3,4} \and {B.~D. Wandelt}\inst{5} \and {K.~M. G\'orski}\inst{5}  
 \and {M. Malaspina}\inst{2} 
\and {M. Bersanelli}\inst{6} 
\and {N. Mandolesi}\inst{2}
\and {A.~J. Banday}\inst{7}
\and {E. Hivon}\inst{8}
} 
\begin{document}
\offprints{maino@sissa.it}
\thesaurus{03.13.2; 12.03.1; 03.19.2}

\institute{
{SISSA, International School for Advanced Studies, Via Beirut 2-4, 
I-34014 Trieste, Italy}
\and
{Istituto TeSRE, Consiglio Nazionale delle Ricerche, Via Gobetti 101, 
I-40129 Bologna, Italy} 
\and
{Dipartimento di Fisica, Universit\'a di Ferrara, Via Paradiso 12, I-44100 
Ferrara, Italy}
\and
{INFN, Sezione di Ferrara, Via Paradiso 12, I-44100 Ferrara, Italy}
\and
{TAC, Theoretical Astrophysics Center, Juliane Maries Vej 30, 
DK-2100 Copenhagen, Denmark}
\and
{IFC, Consiglio Nazionale delle Ricerche, Via Bassini 15, 
I-20133 Milano, Italy}
\and
{MPA, Max Planck Inst. f\"{u}r Astrophysik, Karl-Schwarzschild Str. 1, D-85740,
Garching, Germany}
\and
{CalTech, California Institute of Technology, 1200 East California Boulevard, CA-91125
Pasadena, USA}
}

\maketitle
\markboth{D. Maino et al.: Analysis of $1/f$ noise}
{D. Maino et al.: Analysis of $1/f$ noise}

\begin{abstract}
We study the impact of $1/f$ noise on the {\sc Planck} Low Frequency Instrument
(LFI) observations (Mandolesi \etal 1998) and describe a 
simple method for removing striping effects from the maps for a number of
different scanning strategies. 
A configuration with an angle between telescope optical axis
and spin-axis just less than 90$^\circ$ (namely $\simeq 85^\circ$) 
shows good destriping efficiency for
all receivers in the focal plane, with residual 
noise degradation $< 1-2\%$.
In this configuration,
the full sky coverage can be achieved for each channel separately
with a 5$^\circ$ spin-axis precession to maintain a constant solar
aspect angle.


\end{abstract}

\section{Introduction}

The great success of \COBE-DMR results 
(Smoot \etal 1992, Bennet \etal 1996a, G\'orski \etal 1996)
had a considerable impact on the
scientific community being the first detection of CMB (Cosmic
Microwave Background) anisotropy and 
supporting the gravitational instability scenario for structure formation.
Detailed predictions for the CMB angular power spectrum in different 
cosmological models and the proper signature of 
different cosmological parameters
(e.g. Zaldarriaga \& Seljak 1997) are now quite well established.
A significant number of new experiments, 
both ground-based and balloon-borne, with sensitivity and angular 
resolution better than \COBE\ (e.g. Lasenby \etal 1998 and De~Bernardis
\& Masi 1998) are going to be designed and constructed 
and will deliver good quality data sets in the next few years.
 
It is also clear (Danese \etal 1996) that only a space mission
free from unwanted contamination from ground and Earth atmosphere 
and with a nearly full-sky coverage can fully exploit the gold mine
of cosmological information imprinted in the CMB anisotropy. 
The two space missions MAP (Microwave Anisotropy Probe) 
(see Bennet \etal 1996b) by NASA and {\sc Planck} 
(Mandolesi \etal 1998, Puget \etal 1998) by ESA are planned for the 
next decade, the first to be launched in the year 2001 and the 
second in 2007. These planned observations,
with their wealth of cosmological information,  
will play a determinant role to fully understand the properties
and the evolution of the universe.
In particular, {\sc Planck} is a third-generation CMB space missions designed 
to achieve unprecedented sensitivity and angular resolution
over the entire sky. The extremely wide frequency range
($\simeq 30 \div 900$~GHz) explored by the combination of two 
focal plane arrays, the Low Frequency Instrument (LFI) and
the High Frequency Instrument (HFI), will allow 
to efficiently separate the frequency independent primary cosmological 
information from astrophysical signatures
(e.g. De~Zotti \etal 1999) by taking 
advantage from their proper frequency dependencies. 

As for any CMB experiment,  great attention has to be devoted 
by the two mission teams to all the possible systematic effects.
In the context of the {\sc Planck} mission we have carried out a 
detailed study of one of these effects: the so-called
$1/f$, or low-frequency, noise which is particularly relevant for
the LFI radiometers, leading to stripes in the final maps which
increase the overall noise level and may alter the statistical 
analysis of the anisotropy distribution.
Recent analytical work by Seiffert \etal (1997) has shown the 
dependency of the $1/f$ noise upon the radiometer characteristics 
such as the bandwidth, the noise temperature, 
payload environment temperature and other quantities properly related to
{\sc Planck}-LFI radiometers. 
They can be combined to define a representative parameter,
the ``knee-frequency'' $f_k$, which has to be kept as low as possible
compared with the spinning frequency $f_s$ of the spacecraft.
Janssen \etal (1996) have indeed demonstrated that for
$f_k \gsim f_s$ a degradation in final sensitivity will
result.

In Sect.~\ref{sourcenoise} we briefly summarize the source of
the $1/f$ noise and the dependency of the knee-frequency upon
the radiometer properties. In Sect.~\ref{destriping} we outline
the framework of our simulations, evaluate the possible impact of 
$1/f$ noise and present the technique
for removing the $1/f$ noise in the context of {\sc Planck}-LFI.
The main results both for un-reduced and reduced $1/f$ noise
are presented in Sect.~\ref{results}.
Our main conclusions and implications for the scanning strategy are
discussed in Sect.~\ref{conclusions}.

\section{Source of $1/f$ noise}
\label{sourcenoise}

\begin{figure}[tb]
\vskip 1truecm
\centerline{
\psfig{figure=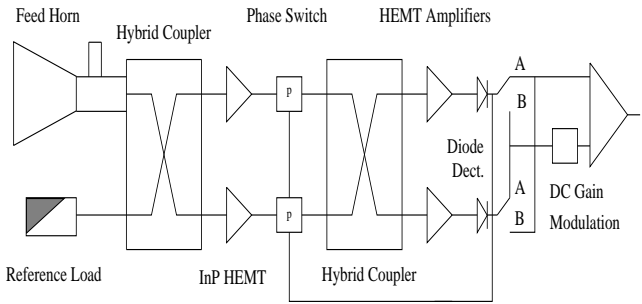,width=8.5cm,height=4cm,angle=-90}}
\caption[]{Schematic representation of the LFI radiometers.}
\label{radiometer}
\end{figure}

The presence of unwanted systematic effects is one of the main
difficulties in CMB anisotropy experiments. 
The so-called $1/f$ noise, typically generated by 
HEMT (High Electron Mobility Transistor) amplifier 
gain instabilities, is
of particular interest for the {\sc Planck}-LFI 
since, if not properly corrected, it may lead to stripes in the final maps
due to the satellite scanning strategy. Therefore it is of great
importance to reduce the impact of such effect both by hardware and
software techniques.  The LFI receiver concept is driven by the need to
reduce instability effects.
Bersanelli \etal (1995) described the design of LFI radiometers 
(Fig.~\ref{radiometer}) which are modified Blum correlation receivers 
(Blum 1959, Colvin 1961) known also as pseudo-correlation or
continuous comparison receivers. The details of the radiometer will
not be described here. The main feature to
note is that the receivers have two inputs: one from one of the two 
polarization directions of a feed-horn and one from a reference load.
The signal is sent by an hybrid coupler in the two ``legs'' of the 
receiver where it is amplified and phase shifted. The signal after a 
second hybrid, is amplified again and detected.
The output signal is proportional to the signal difference at the two 
inputs. This is achieved by differencing successive measurements at 
the detector which, for the receiver design, come from the 
horn and from the load respectively. The ideal situation with respect to
the sensitivity to systematic effects is when the output is nulled. 
It is therefore important that the temperatures of the reference 
load be as close as possible to the $\approx 3$~K temperature of the sky signal. 
Furthermore, a nearly-null output is obtained by 
adjusting properly the DC gain ratio $r$ after the diodes at the value 
$r=(T_x+T_n)/(T_y+T_n)$, where $T_y$ is the 
temperature from the reference load input. 

\begin{figure*}[!th]
\centerline{
\psfig{figure=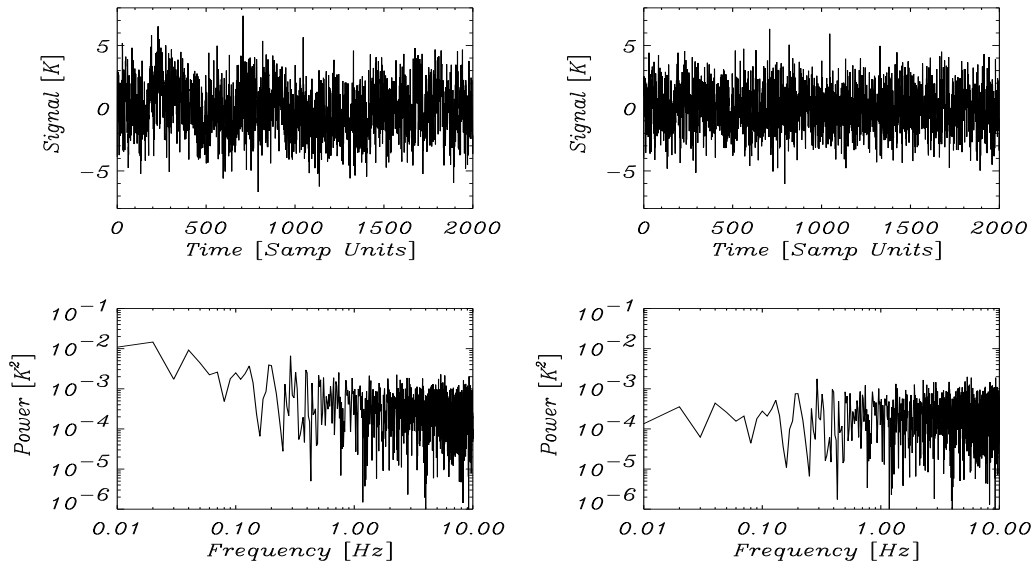,width=14cm,height=8cm}}
\caption[]{Comparison between a pure white noise (right) 
and white+$1/f$ noise (left) in time domain (upper panels) and
and Fourier space (bottom panels). Their difference is more clear in
Fourier space with a flat spectrum for the white noise and the
typical $1/f$ noise spectrum for the low-frequency noise. We used 
a $f_k = 0.8$ Hz for this illustrative case}
\label{example}
\end{figure*}

Seiffert \etal (1997) have reported an accurate
analysis of the source of $1/f$ noise in the final output
of this receiver design.
The ideal white noise sensitivity for this kind of radiometer
is given by:
\begin{equation}
\Delta T_{\rm wn} = \frac{\sqrt{2}(T_n + T_x)}{\sqrt{{B}\tau}} \, ,
\label{whitenoise}
\end{equation}
where $B$ is the bandwidth (for LFI 20\% of the central frequency), $\tau$ is the
integration time, $T_x$ is the signal at horn input and $T_n$ is the noise
temperature of the amplifier.
The main source of $1/f$ noise resulted be gain fluctuations in the HEMT 
amplifiers which have 
a typical $1/f$ spectrum. These fluctuations translate into
fluctuations of the noise temperature $T_n$ with the same spectral shape. 
Noise temperature fluctuations $\Delta T_n$ can mimic 
a change in the radiometer output
\begin{equation}
\Delta T_{\rm equiv}=\sqrt{2}\Delta T_n\left[\frac{1-r}{2}\right]=
\sqrt{2}\left[\frac{A\times T_n}{\sqrt{f}}\right]\left[\frac{1-r}{2}\right] \, ,
\label{equiv}
\end{equation}
where $A$ is the square root of the spectral amplitude of noise 
temperature fluctuations.
Following Seiffert \etal (1997), we define the ``knee-frequency'' when 
$\Delta T_{\rm equiv} = \Delta T_{\rm wn}$. This leads to the expression
\begin{equation}
f_k = \frac{A^2{B}}{8}(1-r)^2\left(\frac{T_n}{T_n+T_x}\right)^2 \, .
\label{knee}
\end{equation}
Typical values are $A\sim 1.8\times 10^{-5}$ for 30 and 44 GHz radiometers and
$A\sim 2.5\times 10^{-5}$ for 70 and 100 GHz. 
Fig.~\ref{example} reports an example of simulated output for a pure white
noise and a white noise plus $1/f$ noise signal: their difference is much more
clear in Fourier space where the $1/f$ noise spectrum shows up at low frequencies. 
For the 30 GHz channel, with $T_n = 9$ K, $T_x = 3$ K, $B$=6 GHz, $T_y = 20$ K
Eq.(\ref{knee}) gives a $f_k \sim 0.046$ Hz. For a 4 K reference load,
which is now the adopted baseline, the knee-frequency reduces to $\approx
1$ mHz. For the 100 GHz channel $f_k$ is higher (0.11 Hz) for a 20 K
reference load and is reduced to less than 1 mHz for the 4 K load.

\section{Scan circles, sky maps and destriping}
\label{destriping}

As reported in Mandolesi \etal (1998), the selected orbit for
{\sc Planck} satellite will be a Lissajous orbit around the L2 Lagrangian point 
of the Sun-Earth system. The spacecraft spins at 1 r.p.m. and the spin
axis is kept on the Ecliptic plane at constant solar angle by 
repointing of 2.5$'$ every hour. The field of view of the two 
instruments is between $\alpha\approx 80\deg-90\deg$ from the spin-axis
direction. 
Hence {\sc Planck} will trace large circles in the sky: these circles 
cross each other in regions close to the Ecliptic poles. The
shape and width of these regions depend upon the angle $\alpha$, the 
scanning strategy and beam location in the focal plane.
The value of the angle $\alpha$ has not yet been fully defined, 
as well as the scanning strategy, which may or may not include a
periodic motion of the spin-axis away from the Ecliptic plane. 
These options depend on a trade-off between different
systematic effects (striping, thermal effects, straylight), which has to
be carefully addressed.

Here we made an attempt in this direction taking into consideration
the $1/f$ noise alone as if all the other systematics were 
negligible.
This is of course an idealised situation but gives us 
an estimate of the impact of this effect that therefore should be regarded
as a lower limit. Nonetheless this result may help for a better design
of the instrument. 
Future work will consider a more complete situation
when a detailed thermal transient analysis will be available for the
Planck spacecraft.

\subsection{The ``flight-simulator''}
\label{flight}
Burigana \etal (1997, 1998) has described in detail the code we have
implemented for the {\sc Planck} scanning strategy and we refer the reader to 
these papers. The relevant geometrical inputs are the beam location in the
focal plane and the angle $\alpha$ between the spin and pointing axis.

For each beam position on the focal plane our code outputs the complete 
data stream. We consider here a reduced version of the actual baseline
for the scanning strategy (actual parameters in parentheses): 
spin-axis shift of 5$'$ every 2 hours (instead
of 2.5$'$ every hour) and three samplings per FWHM of 30$'$ at 30 GHz 
(instead of 12 samplings every 30$'$, i.e. 4 samplings
every 10$'$, the FWHM at 100~GHz; see Mandolesi \etal 1998).
These modifications allow us to explore 
a large region of the parameter space, beam position, $f_k$, scanning
strategy and pointing angle $\alpha$, in reasonable time.
Furthermore we do not consider the single minute data stream
but we take the average over the 120 circles forming a given 2-hours
set. In what follows we run simulations for the  30 GHz channel.

Wright (1996) has shown that possible data filtering on a given 
scan circle may help in reducing the impact of $1/f$ noise. This is
useful for values of $f_k$ typical for ``total power'' receivers which are 
much higher that our 0.05 Hz and we chose  not to include  this
technique here. 

It is clear from Eqs.(\ref{whitenoise}) and (\ref{knee}) that both the sensitivity and
the ``knee-frequency'' depend on the actual temperature in the sky $T_x$
seen by the horn. Our synthetic model for microwave sky emission 
includes  a standard CDM prediction 
for CMB fluctuations plus a model of galactic emission. This model has the
spatial template from the dust emission (Schlegel \etal 1998) but has been
normalized to include contribution from synchrotron, free-free and dust
according to \COBE-DMR results (Kogut \etal 1996). The major foreground
contamination at 30 GHz comes from synchrotron and free-free. We 
then choose to overestimate the overall synchrotron fluctuations by a factor
of $\approx$ 10. This is the worst case scenario with respect to destriping
efficiency (see Sect.~\ref{streamtomap} and \ref{destri}). 

Of course the impact of the sky temperature is small, since $T_x$ is a small 
fraction of the noise temperature $T_n$: this is in fact $\sim$ 10 K 
while $T_x \approx 3$~K at 30~GHz and the maximum galactic contamination adopted here
is about 44 mK.
In this evaluation of $T_x$ we include also a typical environment temperature
of about 1 K.

We do not perform a convolution with a real beam pattern and sky signal
since the real beam pattern will be in general not perfectly symmetric, not
perfectly gaussian
and with sidelobe contributions which 
will introduce other kinds of contamination. Since we want to estimate here only
the impact of $1/f$ noise we convolve our input map with a gaussian
beam with the nominal FWHM (33$'$) of the 30~GHz {\sc Planck}-LFI
channel we consider.

\subsection{Generation of instrumental noise}
\label{gennoise}

We have the possibility to generate different kinds of noise spectra. We work in Fourier 
space and generate the real and imaginary part of Fourier coefficients of our 
noise signal. The noise spectrum has the usual form:
\begin{equation}
S_{noise}(f) = a\left[1 + \left(\frac{f_k}{f}\right)^\beta\right] \, ,
\label{spectrum}
\end{equation}
where $a$ is a normalization factor and $\beta$ takes typical values from
1 to 2.5 depending on the source of noise (gain drift of thermal effects): 
$\beta = 1$ is the $1/f$ noise case.
In general it is a good approximation to take $S_{noise}(f)$ vanishing for
$f<f_{min}$ and $f>f_{max}$.  Delabrouille (1998) proposed 
$f_{min}\sim 1/T_{mission}$ and $f_{max}\sim 1/2T_{samplings}$.

After generating a realisation of 
the real and imaginary part of the Fourier coefficients with 
spectrum defined in Eq.(\ref{spectrum}), we FFT (Fast Fourier Transform;
Cooley \& Tukey 1965;  Heideman \etal 1984) 
them and obtain a real noise stream which has to be normalized to the white 
noise level defined in Eq.(\ref{whitenoise}). 
We use a typical value of $f_k = 0.05$ Hz for the knee-frequency at 30 GHz 
assuming a 20 K load temperature (see Sect.~\ref{sourcenoise}).
We chose to generate noise streams of about $2\times 10^{6}$ data points, 
corresponding to 8 spin axis positions or 16 hours: this seems a 
reasonable compromise among our present knowledge of real hardware behaviour
and the required computational time with respect noise stream length.
The actual time for a generation of a noise stream 16
hours long and one year of mission even in our reduced scanning strategy
will require about 10 hours on a Silicon Graphics machine with 2 Gb of RAM
and clock speed of 225 MHz.
We also verified that the most time consuming operation in our
code is just the FFT $1/f$ noise generation. 

The possibility to generate the noise streams with
a completely arbitrary spectrum in real space 
(e.g. Beccaria \etal 1996 and Cuoco \& Curci 1997 and references therein),
without the use of FFT, seems very promising for considerably reducing 
the computational time (Wandelt \etal 1999, Wandelt \& G\'orski 1999).

\subsection{From data stream to sky maps}
\label{streamtomap}
The final output of our ``flight-simulator'' are 4 matrices with a number
of rows equal to the considered spin-axis positions $n_s$ (for one year
of mission 
$\sim 4320$ in our reduced baseline as in Sect.\ref{flight})
and a number of columns equal to the number of integrations,
weakly dependent on $\alpha$, along one scan circle (here $n_p\sim 2160$). 
The input and output maps are in HEALPix pixelisation scheme (G\'orski
\etal 1998, {\tt http://www.tac.dk/\~{}healpix}); 
we use an input map with a resolution of about 3.5$'$
corresponding to 12 million equal area pixels on the full sky.
  
For each circle, the code outputs are the pixel number ${\bf N}$
at the specified resolutions, the temperature plus total noise 
contribution ${\bf T}$ (white plus $1/f$ noise), the temperature with only 
white noise ${\bf W}$ and the pure signal as observed in absence of
instrumental noise ${\bf G}$.
The {\bf W} and {\bf G} will be used for studying the degradation of $1/f$ 
noise with respect to the ideal pure white noise case and the impact of
scanning strategy geometry on observed pixel temperatures. 

We can arbitrarily choose the temperature output data stream 
resolution from 3.5$'$ to higher values (smaller resolution) which 
set also the output temperature map resolution. 
Regarding the data stream for the pixel number outputs 
we can also use higher resolutions, allowing to test the impact
of using more or less stringent crossing conditions in the destriping
algorithm (see the following section). 

From these data streams it is quite simple to obtain observed simulated maps: 
we make use of ${\bf N}$ and ${\bf T}$ to coadd the temperatures of those 
pixels observed several times during the mission. 
In the same way we build maps with only white noise contribution, without 
receiver noise, as well as a 
sensitivity map knowing how many times a single pixel is observed.

\begin{figure*}[!th]
\centerline{
\psfig{figure=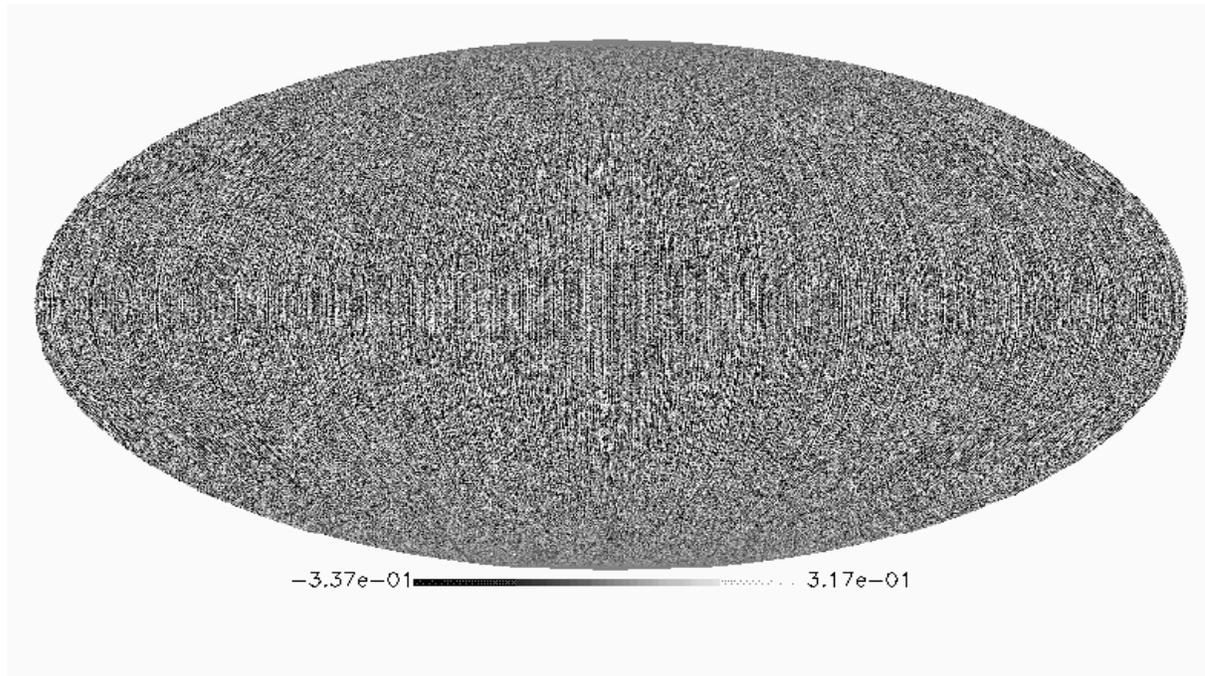,height=9cm,width=16cm,angle=90}}
\caption{Pure noise (white and $1/f$ noise together) map before destriping: stripes are
clearly present. The adopted $f_k$ is 0.05 Hz and units are in mK. See the text for simulation
input parameters.}
\label{maponeonf}
\end{figure*}

In Fig.~\ref{maponeonf} we show a pure noise (white plus $1/f$) map
in Ecliptic coordinates after signal subtraction (${\bf T} - {\bf G}$): 
stripes are clearly present.

\subsection{Destriping techniques}
\label{destri}

We developed a simple technique which is able to eliminate gain drifts due to 
$1/f$ noise. This is derived from the COBRAS/SAMBA Phase A study proposal
(Bersanelli \etal 1996) and from a re-analysis of Delabrouille (1998). 
As reported by Delabrouille (1998) the 
effect of $1/f$ noise can be seen as one or more additive levels,
different for each scan circle. 
We worked with averaged (over 2 hours period) scan circles
and hence we nearly removed drifts within each circle: what is left is related
to the ``mean'' $1/f$ noise level for this observation period.
In fact averaging scan circles into a single ring corresponds to
a low-pass filtering operation. As long as $f_k$ is not far larger than the 
spin frequency, this ensures that only the very lowest frequency
components of the $1/f$ noise survive. Therefore it is a good approximation 
to model the averaged $1/f$
noise as a constant offset $A_i$ for each ring 
for the set of parameters we are using.

We want to obtain 
the baselines for all the circles and re-adjust the signals correspondingly.
Delabrouille (1998) claimed the possibility that working with multiple
baseline per circle ($n_l$ = number of baseline levels)
may result in a better destriping. 
In our code this is equivalent to dividing the
rows of each output matrices in $n_l$ parts: the result will be matrices
with $n_s\times n_l$ rows and $n_p/n_l$ columns.

In order to estimate the different $A_i$ we have to find common pixels
observed by different scan circles and the pixel size in the matrix 
{\bf N} is a key parameter. Increasing the resolution used in {\bf N} 
reduces the number of crossings possibly yielding to lower destriping efficiency,
while adopting resolutions lower than the resolution of the input map 
and of the matrix {\bf T} introduces extra noise, 
related to variations of real sky temperature within the scale corresponding
to the lower resolution adopted in {\bf N}, which is not taken into account
in the estimate of receiver noise and may introduce artifacts in
the destriping code.
The adopted pessimistic galactic emission model, which by construction has 
gradients larger than those inferred by current data, emphasizes this effect
and our simulations are then conservative in this respect.
In the following $N_{il}$, $T_{il}$ and $E_{il}$ will denote the 
pixel number, the temperature and the white noise level  
for the pixel in the $i^{\rm th}$ row and $l^{\rm th}$ column. 
Let us denote a generic pair of different observations of the 
same pixel with an index $\pi$ which will range between 1 and $n_c$, the 
total number of pairs found. In this notation $\pi$ is 
related to two elements of ${\bf N}$: $\pi\rightarrow(il,jm)$ where
$i$ and $j$ identify different scan circles and $l$ and $m$ the
respective position in each of the two circles. 

We want to minimize the quantity:
\begin{eqnarray}
S & = &\sum_{{\rm all\ pairs}} \left[\frac{[(A_i - A_j) -
(T_{il}-T_{jm})]^2}
{E_{il}^2+E_{jm}^2}\right] \nonumber \\ 
  & = & \sum_{\pi=1}^{n_c} \left[\frac{[(A_i - A_j) - (T_{il}-T_{jm})]^2}
{E_{il}^2+E_{jm}^2}\right]_\pi 
\label{firstmini}
\end{eqnarray}
with respect to the unknown levels $A_i$. The sub-index $\pi$ indicates
that each set of $(il,jm)$ is used in that summation. It is clear that $S$ is
quadratic in all the unknown $A_i$ and that only differences between
$A_i$ enter
into Eq.(\ref{firstmini}). Therefore the solution is determined up to
an arbitrary additive constant (with no physical meaning for anisotropy
measurements). We choose then to remove this indetermination by requiring
that $\sum_{h=1}^{n_s} A_h = 0$. This is equivalent to replace
Eq.(\ref{firstmini}) with $S' = S + \left(\sum_{h=1}^{n+s} A_h\right)^2$. 
After some algebra we finally get:
\begin{eqnarray}
\frac{1}{2}\frac{\partial S'}{\partial A_k} & = & \sum_{\pi=1}^{n_c}
\left[\frac{ [(A_i-A_j)-(T_{il}-T_{jm})]\cdot [\delta_{ik}-\delta_{jk}]}
            {E_{il}^2-E_{jm}^2}\right]_\pi \nonumber \\  
           & + & \sum_{h=1}^{n_s} A_h = 0
\label{secondmini}
\end{eqnarray}
for all the $k=1,...,n_s$ (here $\delta$ is the usual Kronecker symbol).
This translates into a set of $n_s$ linear equation
\begin{equation}
\sum_{h=1}^{n_s} C_{kh}A_{h} = B_k, \ \ k=1,...,n_s
\label{thirdmini}
\end{equation}
which can be easily solved. Furthermore we note that by construction 
the matrix ${\bf C}$ of $C_{kh}$ coefficients is symmetric, positive
defined and non singular. The first property permits to hold in 
memory only one half of the matrix ${\bf C}$ (e.g. the upper-right part), 
the second allows us to solve the linear system without having to
exchange rows or columns (Strang 1976), so preserving the symmetry.
The non-singularity of ${\bf C}$ is true provided that there are
enough intersections between different circles and hence is related
to the resolution at which we look for common pixels between different
scan circles, to the scanning strategy and beam location.
A detailed discussion of numerical 
algorithm for solving this system with significant saving of 
required RAM is presented in Burigana \etal (1997).

It is interesting to note that the applicability of 
this destriping technique does not depend upon
any a-priori assumption about the real value of $f_k$ or the
real noise spectral shape since it can work also for different
values of the exponent $\beta$ in Eq.(\ref{spectrum}), althought 
its efficiency could be in principle partially optimized in relation
to the properties of the noise and of the sky gradients.

\section{Simulations results and scanning strategy}
\label{results}

We first consider an angle $\alpha=90^\circ$ between telescope 
and spin axes and a beam location
with ($\theta_B,\phi_B)=(2.8^\circ,45^\circ$). Here $\theta_B$ is the 
angle from the optical axis: this is a typical value for the 100 GHz
(with a $\simeq 1.5$~m aperture off-axis Gregorian telescope)
horns while the 30 GHz beams are placed at larger $\theta_B$ values. This
assumption is therefore conservative with respect to the destriping efficiency,
since in this case the region of crossings between scan circles is smaller 
and closer to Ecliptic poles. The angle $\phi_B$ is the beam center
displacement from the axis given by the intersection between the sky field of
view plane and the plane containing the telescope and spin axes (see Burigana
\etal 1998). Our choice of $\phi_B=45^\circ$ is intermediate
between 0$^\circ$ (or 180$^\circ$) and 90$^\circ$ which are equivalent for the destriping
respectively to
an on-axis beam (with crossings only at Ecliptic poles for $\alpha=90^\circ$)
and a beam which spreads crossings over the wider possible region. 
The adopted beam location is therefore a non-degenerate 
case although non optimal.  

The knee frequency $f_k=0.05$ Hz and the microwave sky 
emission is modeled as in Sect.\ref{flight}.

\begin{figure}[!t]
\centerline{
\hskip -0.5truecm
\psfig{figure=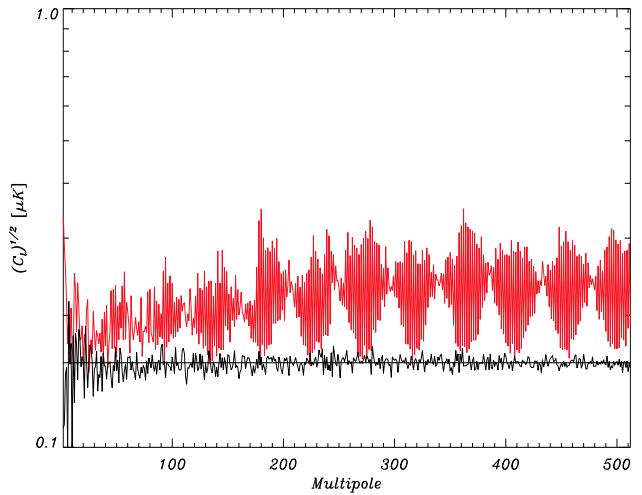,height=7cm,width=9cm}}
\caption[]{Noise power spectra at 30GHz before destriping. 
Simulations parameters are: $\alpha=90\deg$, $(\theta_B,\phi_B)=(2.8\deg,45\deg)$, 
$f_k=0.05$ Hz and spin-axis always on the Ecliptic plane. Theoretical white noise
level is also reported for comparison. The excess of noise is about 40\% over the
white noise level.} 
\label{eclbef}
\end{figure}

We evaluated the impact of $1/f$ noise 
both in terms of added $rms$ noise and of angular power spectrum. 
The output of our simulations is expressed in antenna
temperature and has to be converted into thermodynamic temperature 
for the comparison with CMB power spectrum, as usually predicted in 
theoretical models.
At 30 GHz the conversion factor is very close to unity ($rms_{th} \simeq
rms_{A}$, where $rms_A$ is the $rms$ in antenna temperature).

\begin{figure*}[!th]
\centerline{
\psfig{figure=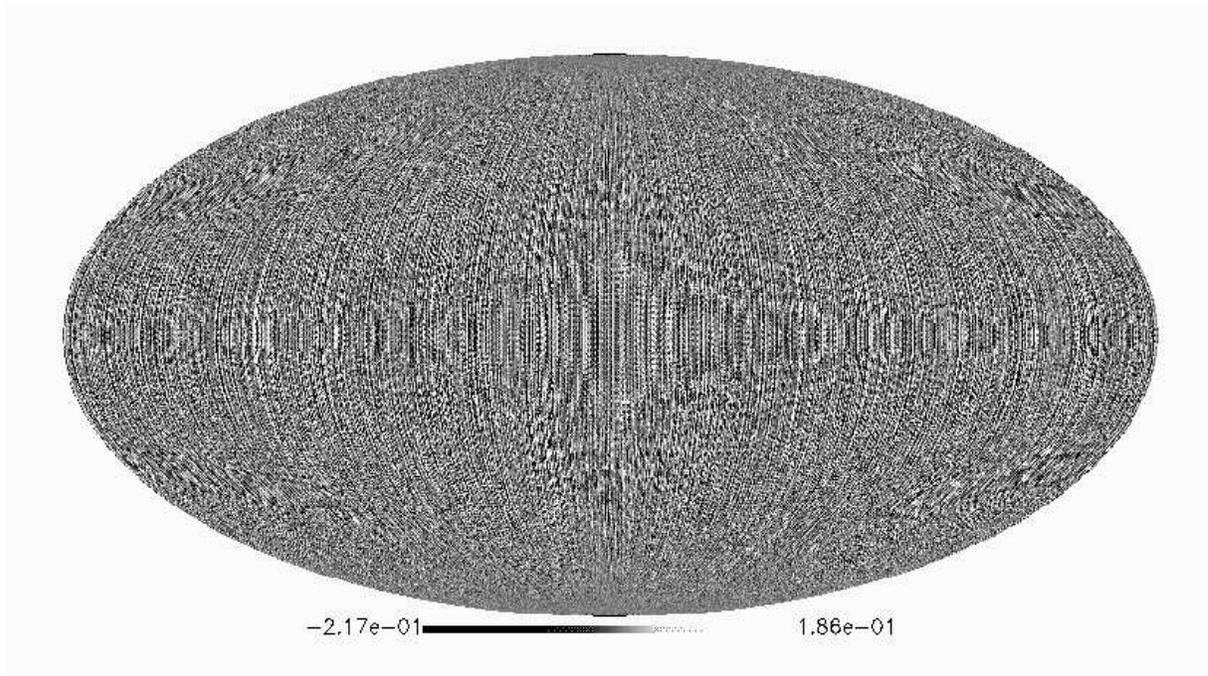,height=9cm,width=16cm,angle=90}}
\caption[]{Baselines recovered from destriping algorithm for the case reported
in Fig~\ref{maponeonf}. }
\label{baselines}
\end{figure*}
\begin{figure*}[!th]
\centerline{
\psfig{figure=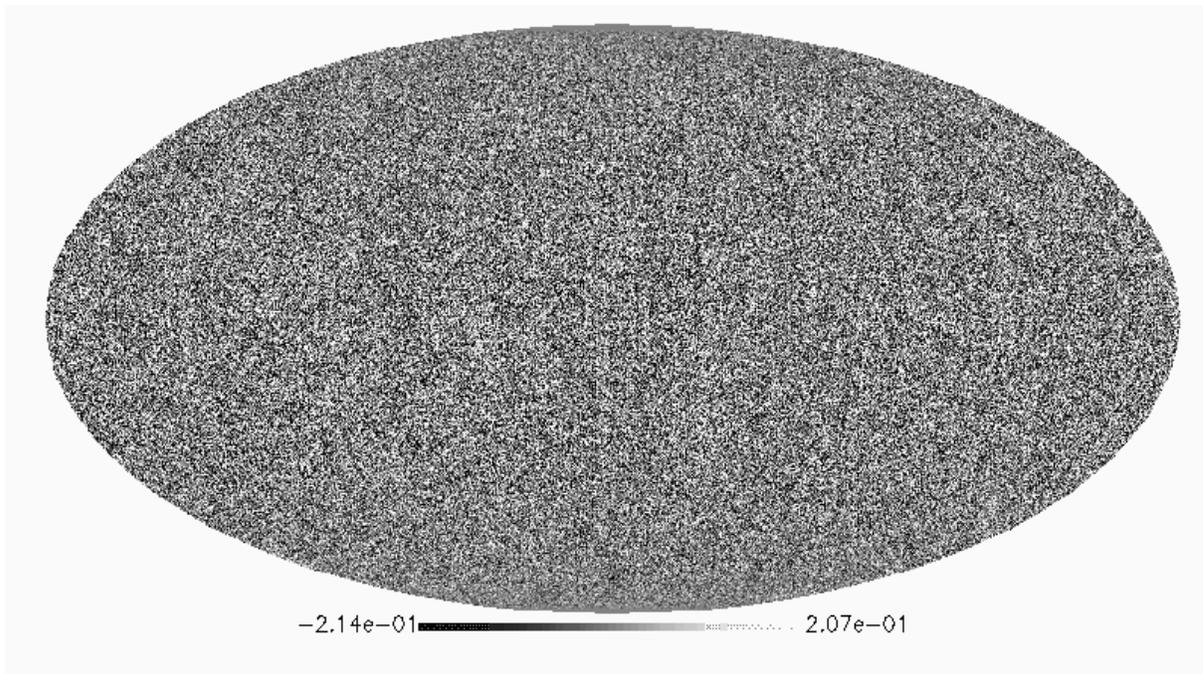,height=9cm,width=16cm,angle=90}}
\caption[]{Noise map after destriping for the same case illustrated in Fig.~\ref{maponeonf}: stripes are no more visible.}
\label{nostripes}
\end{figure*}
The white noise power spectrum can be derived analytically knowing the 
total number of pixels in the sky and pixel sensitivity :
\begin{equation}
C_{\ell,wn} = \frac{4\pi}{N_{\rm pix}^2}\sum_{i=1}^{N_{\rm pix}} \sigma_i^2 
             = \frac{4\pi}{N_{\rm pix}} \langle \sigma^2 \rangle \, .
\label{clwhitenoise}
\end{equation}

\subsection{Morphology of the $C_\ell$ artefacts due to $1/f$ noise}

We now give a  short theoretical argument for the expected morphology of the
$C_\ell$ power spectra
of the simulated striping pattern.  For scanning without wobbling or
precession, 
 two angular scales fully determine the structure
of the striping pattern on the sky. One is the half
opening angle $\lambda_1=90^\circ$ of the ecliptic equator which
the spin axis traces on the sky as {\sc Planck}
completes a full orbit about the
sun.
This makes the noise pattern symmetric under parity, on
average. It follows that the
correlation function for the noise pattern is 
symmetric about $90^\circ$, with a strong peak at $180^\circ$  and
for the $1/f$ component  $C_\ell=0$ for all odd $\ell$.

The other scale is the half opening angle of the scanning rings
 $\lambda_2\simeq \alpha - \theta_B\sin(\phi_B) ,
\; \theta_B\ll \pi/2$. This leads to
peaks in the  correlation function at 
$2\lambda_2$ and by symmetry at $180^\circ-2\lambda_2$.

The combination of these two nearby scales leads to beats in the   
$C_\ell$, which are visually apparent 
as the ``blob'' in Figs.  4,8,10,12 and 14. 
Treating this effect in the same way as the appearance of
fringes in an interference pattern
we can calculate the width in $\ell$ of each blob as
$\Delta \ell = 90/\left(\lambda_2-\lambda_1\right)$. 
This is in
quantitative agreement with our numerically computed spectra. It also
explains the absence of beats in Figure 10, where
$\lambda_2=\lambda_1=90^\circ$ and the power spectrum can be
understood as a single blob with infinite $\Delta \ell$.

For scanning strategies with some kind of ``wobble'', ie.~where the
spin axis does not trace out the ecliptic equator on the sky, 
the correlation function will still be
approximately symmetric if the amplitude of the wobble is small compared to
$90^\circ$ and the above statements remain approximately true. However, the
fact that the symmetry is weakly broken can be seen in Figure 8, where
the interference effect is slightly washed out and the beats  do not
reach all the way down to the white noise level.

The offset  of the blobs compared to the white noise level 
is given by the $rms$ excess power due to
the $1/f$ noise component.
At first sight it may be surprising that other than this normalisation,
the shape of the excess power 
power spectrum $C_\ell$ of the 
striping pattern is independent of the spectral content of the noise, as
parametrised by the knee frequency. This is  due to the filtering
effect of co-adding scan circles into rings discussed above. 
In the regime which our
simulations probe, the structure which remains
on each ring is well described by a constant offset plus white    
noise. Hence, as long as sufficiently many scan circles are co-added,
changing the knee frequency simply scales the relative contribution of
$1/f$ and white noise to the noise pattern $C_\ell$ .

\subsection{Simulation results}

\begin{figure}[!th]
\hskip -0.5truecm
\psfig{figure=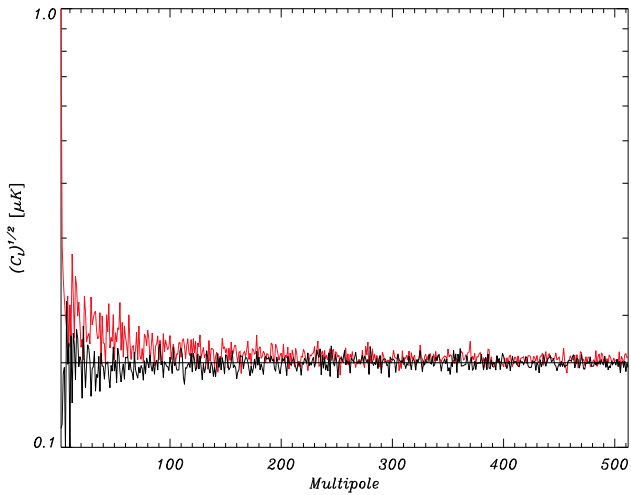,width=9cm,height=7cm}
\caption[]{Noise power spectra at 30GHz after destriping. Simulation parameters
are the same of Fig.~\ref{eclbef}. Now the added noise is only 1-2\% of the
white noise level. White noise spectra are also reported for comparison.}
\label{eclaft}
\end{figure}

In Fig.~\ref{eclbef} we show the square root of the power spectrum (which
is roughly proportional to $rms$ contribution of temperature per $\ell$-bin) 
before applying the destriping
technique: the solid line is the white noise level as derived from 
Eq.(\ref{clwhitenoise}) and the superimposed spectrum is derived from
a simulation with only white noise included. The agreement is very good
which confirms the accuracy of our simulator and map-making algorithm.
The gray line is the global noise spectrum (white and $1/f$ noise together).
The spectrum is clearly non-white: blobs are present. The excess of noise in terms of
both $rms_{th}$ and $\sqrt{C_\ell}$ is about 40\% of white noise level. 
Fig.~\ref{nostripes} shows the noise map after applying our destriping code:
stripes are no more evident and Fig.~\ref{baselines} shows the
recovered baselines by destriping algorithm. This situation can be 
quantified by
computing the noise power spectrum of the destripped map, as shown 
in Fig.~\ref{eclaft}. Now the spectrum is considerably flatter 
and no more blobs are present, with an overall noise excess
of 1-2\% over the white noise level. This confirms the efficiency of 
the destriping algorithm, under the above mentioned simplifying
assumptions.
On the other hand it is clear from Fig.~\ref{eclaft} that at large scales,
corresponding to $\ell\lsim 100$, a non-negligible residual 
contribution is present in the noise spectrum. A lower knee 
frequency (achievable with the reference loads at 4K) is necessary to 
further reduce the effect.
We also investigated if different scanning strategies may help in destriping
efficiency.
Possible periodic motion of the
spin-axis away from the Ecliptic plane have the effect of broadening the
region of crossings between different circles. This goes in the direction
of removing possible degeneracies in the destriping system.
We implemented both sinusoidal oscillations and precessions: the first does not
preserve the spacecraft solar illumination and this is likely to induce thermal effects 
and drifts in the data.
\begin{figure}[!th]
\hskip -0.5truecm
\psfig{figure=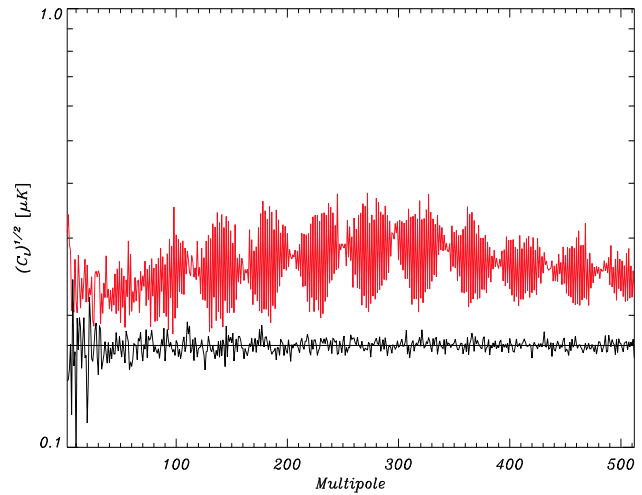,width=9cm,height=7cm}
\caption[]{Noise power spectra at 30GHz Before destriping. Simulation parameters
are the same of Fig.~\ref{eclbef} except for the scanning strategy which includes
here precession (see the text). The added noise is $\sim 40$\% of the
white noise level. White noise spectra are also reported for comparison.}
\label{prebef}
\end{figure}
\begin{figure}[!th]
\hskip -0.5truecm
\psfig{figure=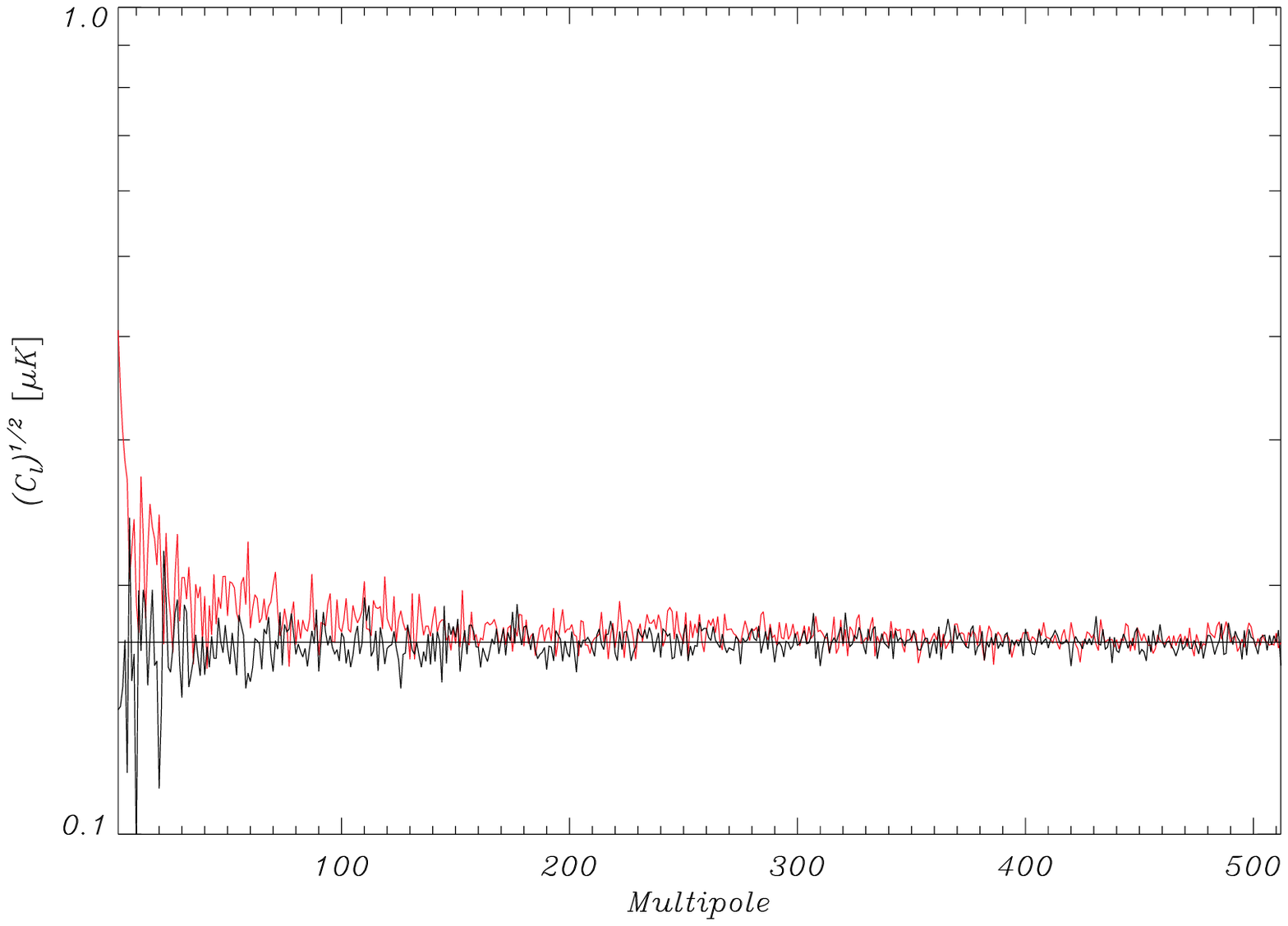,width=9cm,height=7cm}
\caption[]{Noise power spectra at 30GHz after destriping. 
Simulation parameters are the same of Fig.~\ref{prebef}. The added noise is now only 1-2\%
of the white noise level. White noise spectra are also reported for comparison.}
\label{preaft}
\end{figure}

The second is then preferable, since it keeps the solar angle
constant as the satellite moves the spin-axis.
For both the motions we performed 10 complete oscillations per year of mission
with 10\deg\  amplitude.
The results before and after destriping are nearly the same for the two cases
and Fig.~\ref{prebef}
and Fig.~\ref{preaft} report noise power spectra for the precession motion: before
destriping blobs are still present and disappear after destriping. The excess of
noise is $\sim$ 40\% and 1-2\% before and after destriping respectively. 

\begin{figure}[!th]
\hskip -0.5truecm
\psfig{figure=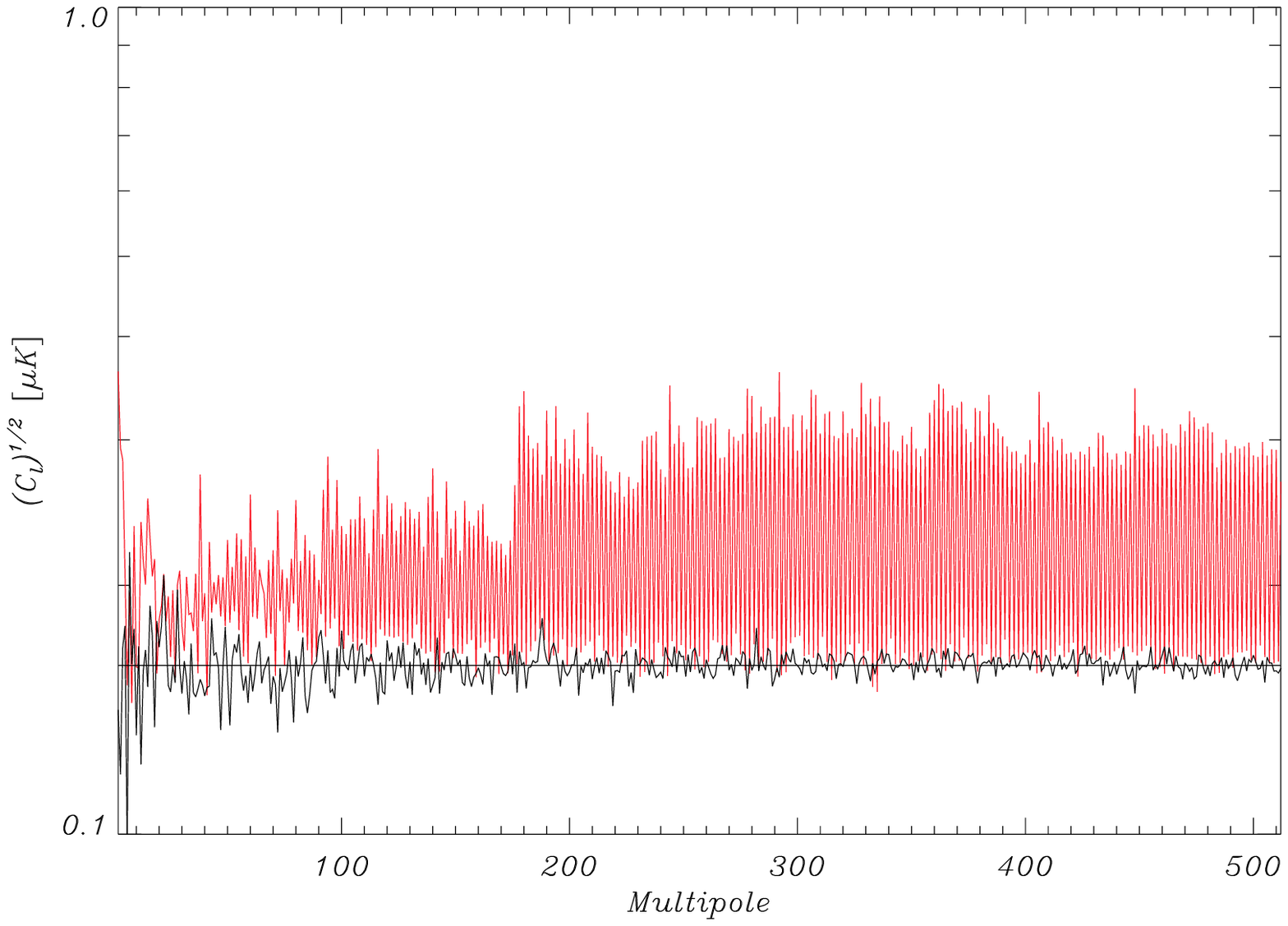,width=9cm,height=7cm}
\caption[]{Noise power spectra at 30GHz Before destriping. Simulation parameters
are the same of Fig.~\ref{eclbef} but the beam position is now 
$(\theta_B,\phi_B)=(0\deg,0\deg)$. The added noise is $\sim 42$\% of the
white noise level. White noise spectra are also reported for comparison.}
\label{onbef}
\end{figure}
\begin{figure}[!th]
\hskip -0.5truecm
\psfig{figure=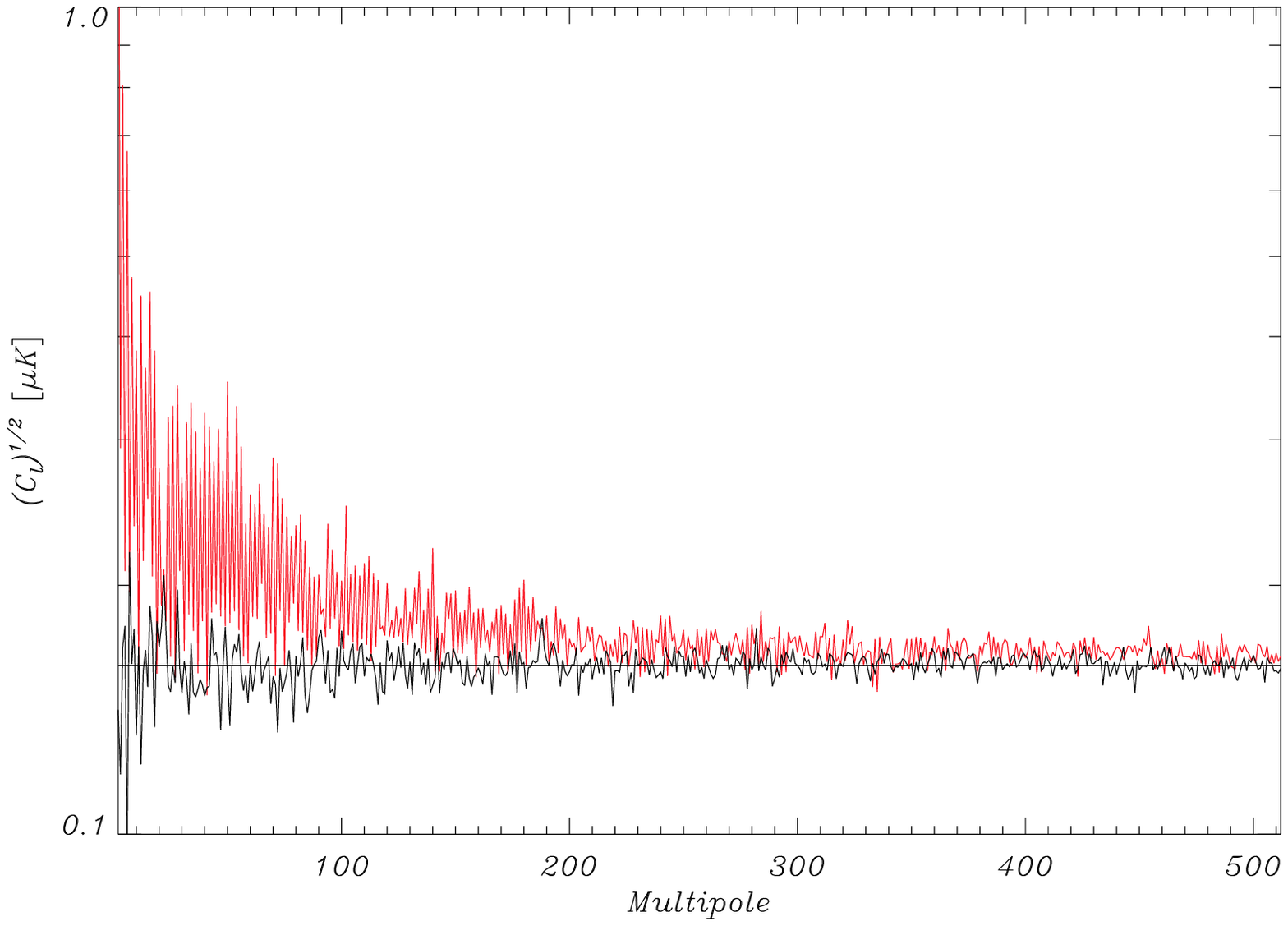,width=9cm,height=7cm}
\caption[]{Noise power spectra at 30GHz after destriping. 
Simulation parameters are the same of Fig.~\ref{onbef}. The added noise is now 
larger than previous cases, being 5\% of the white noise level. 
White noise spectra are also reported for comparison.}
\label{onaft}
\end{figure}
We consider also a case of an on-axis beam $(\theta_B,\phi_B)=(0\deg,0\deg)$
with $\alpha=90\deg$. This situation is representative of 
horns which, in the focal plane arrangement, are
placed close to the scanning direction: from the destriping point of view
these horns are indeed equivalent to on-axis horns
and intersections between different scan circles are nearly only at Ecliptic
poles. This represents a degenerate situation with respect to destriping
efficiency. Fig.~\ref{onbef} and Fig.~\ref{onaft} show noise power
spectra for this on-axis case
before and after destriping respectively: the geometry of the
simulation is changed and the same for blob shape which is not well
defined. Here the level of added noise is $\sim 42\%$. After destriping
we are left with an excess of noise of the order of 5\% of the white
noise and the noise spectrum is considerable less flat than before.
Such excess noise and a non-flat
spectrum around $\ell \sim 100-200$ (where the first CMB Doppler peak
is expected) after destriping is clearly not acceptable for 
the feeds aligned with the scan direction. Moving the spin axis 
is a way to remove this degeneracy, as well as to complete the sky
coverage for all channels.
\begin{figure}[!th]
\hskip -0.5truecm
\psfig{figure=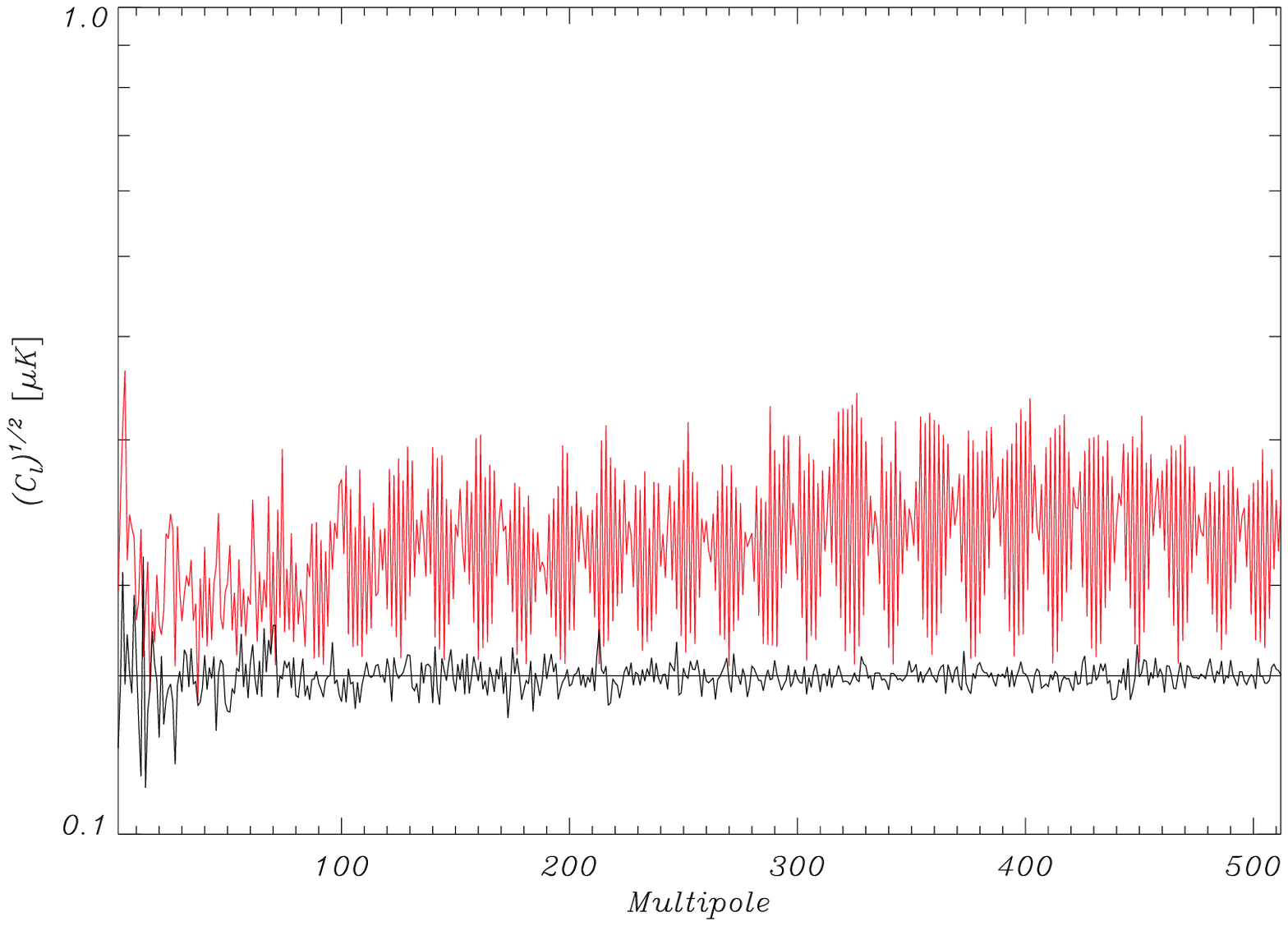,width=9cm,height=7cm}
\caption[]{Noise power spectra at 30GHz Before destriping. Simulation parameters
are the same of Fig.~\ref{onbef} but the angle $\alpha$ is now 
set to 85\deg. The added noise is $\sim 42$\% of the
white noise level. White noise spectra are also reported for comparison.}
\label{85bef}
\end{figure}
\begin{figure}[!th]
\hskip -0.5truecm
\psfig{figure=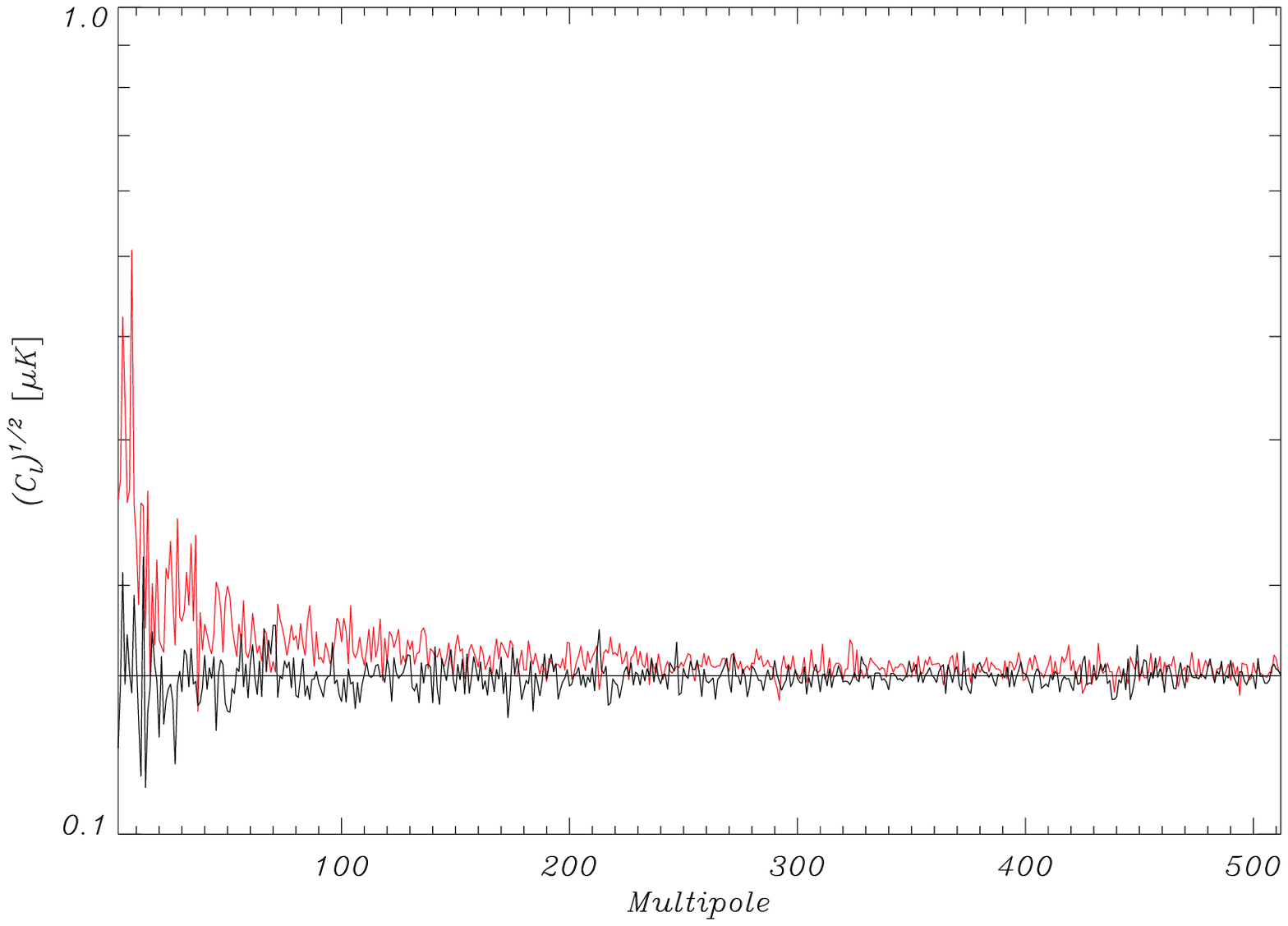,width=9cm,height=7cm}
\caption[]{Noise power spectra at 30GHz after destriping. 
Simulation parameters are the same of Fig.~\ref{85bef}. The added noise is now 
only 2\% of the white noise level. White noise spectra are also reported
for comparison.}
\label{85aft}
\end{figure}
Keeping fixed the beam position we move the angle $\alpha$ from 90\deg\ to
85\deg: this of course leaves small regions around Ecliptic poles which are
not observed but has the advantage to enlarge the scan circle crossing region.
Fig.~\ref{85bef} and Fig.~\ref{85aft} show noise power spectra before and
after destriping respectively for this 85\deg, on-axis case: now blobs
are clearly visible with dimensions in $\ell$-space different from Fig.~\ref{eclbef}
due to different geometrical configuration. These blobs are 
completely removed after destriping leaving a level of added noise after
destriping of only $\lsim$ 2\% over the white noise. 
This is essentially what we obtained for the off-axis configuration.
\begin{figure}[!th]
\hskip -0.5truecm
\psfig{figure=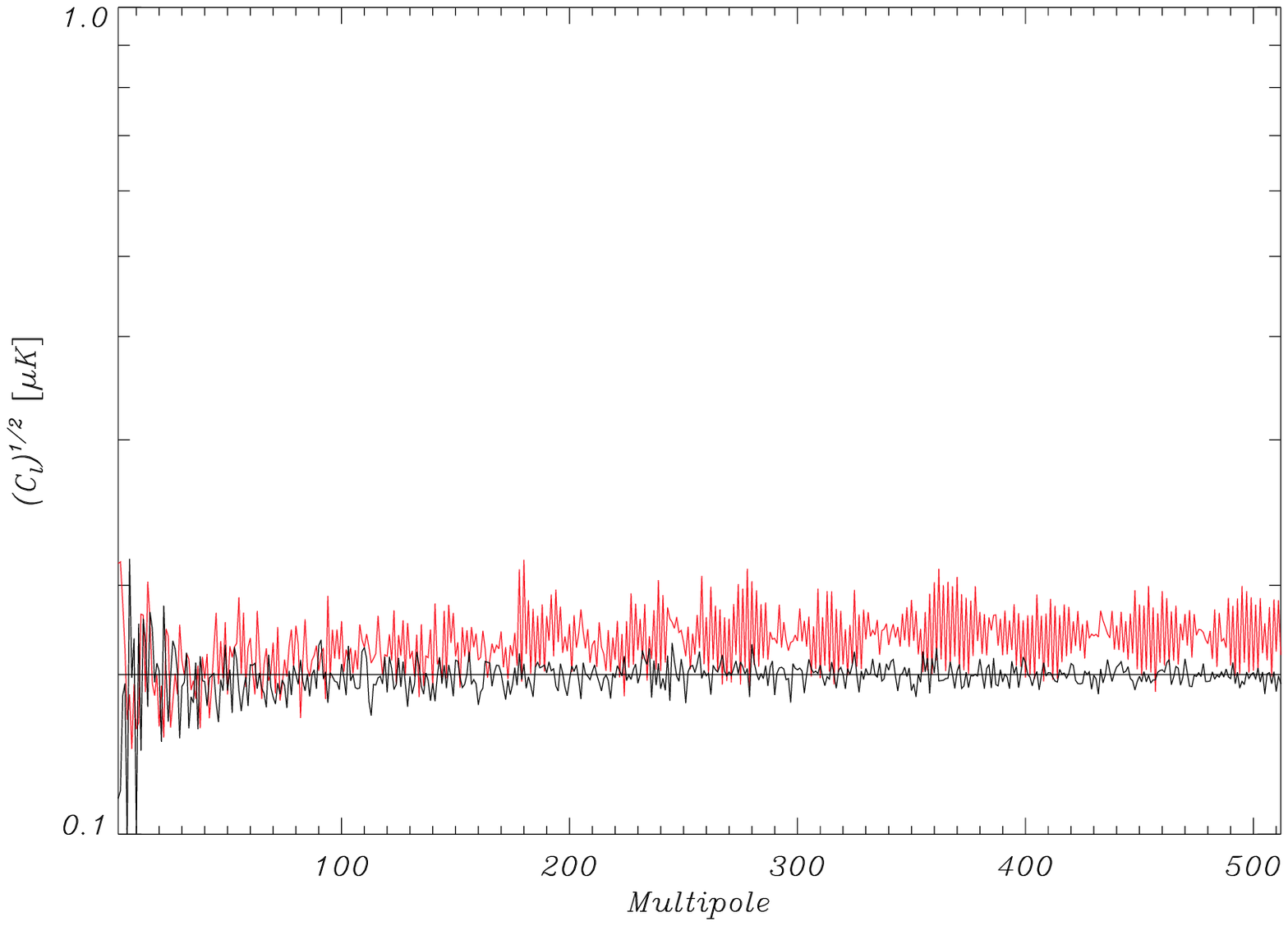,width=9cm,height=7cm}
\caption[]{Noise power spectra at 30GHz before destriping. 
Simulation parameters are the same of Fig.~\ref{eclbef} but now
$f_k=0.01$ Hz. The added noise is now only 9\%
of the white noise level. White noise spectra are also reported for comparison.}
\label{001bef}
\end{figure}
All the simulations considered so far were performed with a $f_k = 0.05$
Hz which is derived from Eq.(\ref{knee}) assuming a $T_y=20$ K of
reference load. 
To decrease the residual effect, a 
4 K reference load is being designed, exploiting the 4~K stage of the
HFI. In this configuration it is expected that the
resulting knee frequency would be less than 10 mHz. We have run a
simulation with $f_k = 0.01$~Hz with the usual off-axis configuration with
$\alpha=90\deg$.  As for the other cases we report noise power spectra
before and after destriping in Fig.~\ref{001bef} and Fig.~\ref{001aft}. 
\begin{figure}[!th]
\hskip -0.5truecm
\psfig{figure=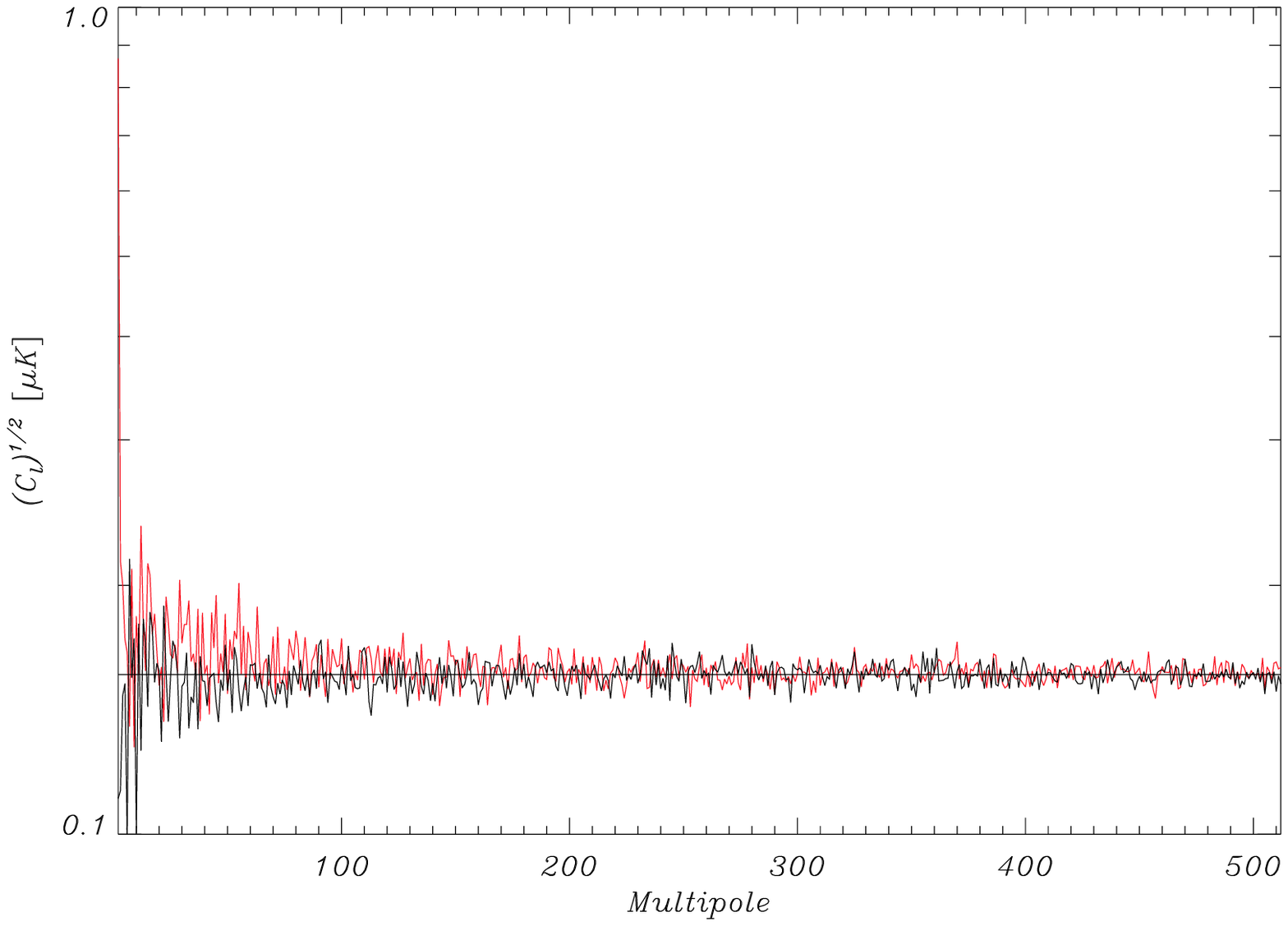,width=9cm,height=7cm}
\caption[]{Noise power spectra at 30GHz after destriping. 
Simulation parameters are the same of Fig.~\ref{001bef}. The added noise is now 
only 0.5\% larger the white noise level. White noise spectra are also
reported for comparison.}
\label{001aft}
\end{figure}
It is interesting to note that now the level of added noise is 9\% before
destriping and reduces only to 0.5\% after applying destriping algorithm.

Although a comparison among different kinds of systematic effects
(Burigana \etal 1999) and studying how they combine in LFI observations is 
out of the aim of the present work, we have tried a first attempt 
in considering the 
sensitivity of this destriping algorithm to one of the 
most relevant source of contamination in Planck measurements, 
the beam shape asymmetry.
We have considered two simulations: one with a pure circular
gaussian beam and another with a highly elliptical gaussian beam
with a ratio of 1.7 between major and minor axis, that exibit 
rms differences of about 5~$\mu$K in the temperatures observed 
in the same sky directions (Burigana \etal 1998). 
The impact of beam ellipticity on
destriping efficiency appears to be small: it has a maximum effect
which is $\sim$3 times smaller than the white noise level between 
$100<\ell<200$ (Burigana \etal 1999). 
This preliminary result suggests that the adopted algorithm may be only 
weakly dependent on the contamination introduced by other
kinds of systematic effects; further investigations are demanded 
to confirm this perspective.

\section{Discussion and conclusion}
\label{conclusions}

We have reported here an extensive study of the $1/f$ noise contamination
for different scanning strategies, beam location on the 
focal plane as well as for different values of the knee frequency.
Even in this idealised situation (all other systematics are well
under control) we
can gain useful information from our study.

First of all from these simulations it seems that moving the spin-axis 
away from the ecliptic plane
does not significantly help the destriping efficiency for
typical LFI beam locations and, concerning the $1/f$ noise 
alone as source of drifts,
it would be preferable to keep the spin-axis always on the Ecliptic plane.
This is clear when we compare results as those in Fig~\ref{eclaft} and
Fig.~\ref{preaft} which are practically identical. 

Furthermore for most of the LFI beams the choice of $\alpha=90\deg$ would
be acceptable: the destriping left only $\lsim 2\%$ of excess noise with
respect to the pure white noise case. On the other hand, some LFI beams are
equivalent to on-axis beam which, from Fig.~\ref{onaft}, is clearly a
degenerate case. A smaller value of $\alpha$ (namely 85\deg) breaks this
degeneracy at an acceptable level yielding the usual destriping
efficiency. From these two points an indication of a possible choice of
the scanning strategy and instrument configuration arises: with
$\alpha=85\deg$ and precession of the spin-axis (no thermal drifts) with
only 5\deg amplitude (half of what we considered here) 
appears satisfactory for de-striping performances while preserving
full-sky coverage for all channels. This allows data redundancy but
introduces irregularities in the integration time distribution,
which may be an issue for the data analysis. Without spin axis
modulations, a quite complete sky coverage and a smooth integration 
time distribution at each frequency can be achieved only by assembling 
data from different receivers, losing redundancy. 

We have considered also the destriping efficiency when using more than a 
single baseline level $A_i$ per scan circle: results reported in
Burigana \etal (1997) indicate that for the $1/f$ noise case no 
significant improvements are achieved:
either the $1/f$ noise level per circle is almost constant and equivalent to 
a single baseline, or the accuracy in the determination of more levels
per  circle decreases.
The whole set of simulations seems to indicate that there is enough redundancy
of observations to remove at acceptable level the contamination due to
$1/f$ noise even if we require a more strict condition of crossings between
scan circles.
Of course, the performance of this destriping code could be partially
optimized in the future by appropriately choosing the number of levels 
per circle and the crossing condition according to the dominant 
kind of instrumental noise (the parameters $f_k$ and $\beta$),
the magnitude of the gradients in the sky emission and our
knowledge of other contamination sources.

For what concerns properly the $1/f$ noise, an important 
indication comes from the simulation with 
$f_k=0.01$~Hz: the excess of noise before destriping reduces by a 
factor $\simeq 4$ with respect to the case $f_k=0.05$~Hz. Reaching
$f_k=0.01$~Hz is a goal of the LFI design, which will ensure
a clean and robust observation.
In addition the extra noise level after destriping decreases, at
least under these ideal assumptions, by a factor $\simeq 3$. 

There are many open issues both astrophysical and instrumental. Regarding the 
first, the microwave emission model we use, although pessimistic for
what concerns galactic emission, can be
completed with the inclusion of different foreground contributions. 
The emission from extragalactic point sources and in particular
their variability may decrease our destriping efficiency.
We have also not considered here any other source of possible systematic
effects such as thermal effects and stray-light contamination induced by
Galaxy emission. Burigana \etal (1999) describe in detail the impact of
stray-light contamination.  Also these effects may in principle
degrade the accuracy in removing $1/f$ noise stripes. 

As proved by the scientific experience in many years of work 
both in physics and in cosmology and astrophysics, efficient data 
analysis tools can significantly improve the quality of the information
extracted from the data, provided that the systematic effects are well
understood, but the first and most important step in projecting
experiments is to reduce all the contaminations at the lowest possible levels.
It is then of great importance to decrease as much as possible the impact
of $1/f$ noise before destriping and $f_k = 0.01$ Hz is an important
goal for instrument studies and prototypes.

\begin{acknowledgements}
We gratefully acknowledge stimulating and helpful discussions with 
L. Danese, J. Delabrouille and M. Seiffert.
\end{acknowledgements}

\noindent
{\bf References}
\rref{Beccaria, M., \etal, 1996, ``Gravitational wave search: real time 
data analysis strategies on paralle computers'', NTS-96024}
\rref{Bennet, C. \etal, 1996a, ApJ, 464, L1}
\rref{Bennet, C. \etal, 1996b, Amer. Astro. Soc. Meet., 88.05}
\rref{Bersanelli, M. \etal, 1995, Int. Rep. TeSRE/CNR 177/1995}
\rref{Bersanelli, M. \etal, 1996, ESA, COBRAS/SAMBA Report on the Phase A 
Study, D/SCI(96)3}
\rref{Blum, E.J., 1959, Annales d'Astrophysique, 22-2, 140}
\rref{Burigana, C. \etal, 1997, Int. Rep. TeSRE/CNR 198/1997}
\rref{Burigana, C. \etal, 1998, A\&ASS, 130, 551}
\rref{Burigana, C. \etal, 1999, A\&A, to be submitted}
\rref{Colvin, R.S, 1961, Ph.D. Thesis, Stanford University}
\rref{Coolet, J.W. \& Tukey, J.W., 1965, ``An algorithm for machine
calculation of complex Fourier series'', Math. Computat., vol. 19, 297} 
\rref{Cuoco, E., \& Curci, G., 1997, ``Modeling a VIRGO-like noise spectrum - Note 1'',
VIR-NOT-PIS-1390-095, 14/07/97}
\rref{Danese, L. \etal, 1996, Astro. Lett. Comm., 35, 257}
\rref{De~Bernardis, P. \& Masi, S., 1998, Proceedings of the XXXIII$^{rd}$
Rencontres de Moriond, pg. 209}
\rref{Delabrouille, J., 1998, A\&ASS, 127, 555}
\rref{De~Zotti, G. \etal, 1999,  to appear in Proceedings of the Conference
``3 K Cosmology'', Roma, Italy, 5-10 October 1998, 
AIP Conference Proc, astro-ph/9902103}
\rref{G\'orski, K.~M. \etal, 1996, ApJ, 464, L11}
\rref{G\'orski, K.~M., Hivon, E., \& Wandelt, B.~D., 1998, 
to appear in ``Proceedings of the MPA/ESO Conference on
     Evolution of Large-Scale Structure: from Recombination to
Garching'',  Banday, A.~J. \etal (Eds.), astro-ph/9812350 }
\rref{Heideman, M.T., Johnson, D.H. \& Burrus, C.S., 1984, ``Gauss and
the history of the FFT'', IEEE Acoustics, Speech and Signal Processing
Magazine, Vol.1, 14}
\rref{Janssen, M. \etal, 1996, astro-ph/9602009}
\rref{Kogut, A. \etal, 1996, ApJ, 460, 1}
\rref{Lasenby, A.~N., Jones, A.~W., \& Dabrowski, Y., 1998, 
Proceedings of the XXXIII$^{rd}$ Rencontres de Moriond, pg. 221}
\rref{Mandolesi, N. \etal, 1998, Planck LFI, A Proposal Submitted to the
ESA}
\rref{Puget, J.~L. \etal, 1998, HFI for the Planck Mission, A Proposal
Submitted to the ESA}
\rref{Schlegel, D.~J. \etal, 1998, ApJ, 500, 525}
\rref{Seiffert, M. \etal, 1997, Rev. Sci. Instrum, submitted}
\rref{Smoot, G.F. \etal, 1992, ApJ 396, L1}
\rref{Strang, G., 1976, ``Linear Algebra and Its Applications'', Academic
Press, Inc.}
\rref{Wandelt, B.~D. \& G\'orski, K.~M., 1999, in preparation}
\rref{Wandelt, B.~D. \etal, 1999, Communication at the LFI Consortium
Meeting, Firenze, 25-26 March 1999}
\rref{Wright, E.~L., 1996,  in IAS CMB
Data Analysis Workshop, Princeton, 22 Nov 1996, astro-ph/9612006 }
\rref{Zaldarriaga, M. \& Seljak, U., 1997, Phys. Rev. D., 55, 7368}
\end{document}